\documentclass{article}

\usepackage{amstext,amsmath,amssymb}
\usepackage[dvips]{color}
\usepackage{graphicx}
\usepackage{epsfig}
\usepackage{cite}

\begin{document}

\setcounter{secnumdepth}{1}

\bibliographystyle{unsrt}

\title{A physically meaningful method for the comparison of potential energy functions}

\author{Jos\'e Luis Alonso$^{1,2}$ and Pablo Echenique$^{1,2}$
\vspace{0.4cm}\\ $^{1}$ {\small Instituto de Biocomputaci\'on y
F\'{\i}sica de los Sistemas Complejos (BIFI),}\\ {\small Edificio
Cervantes, Corona de Arag\'on 42, 50009, Zaragoza, Spain.}\\ $^{2}$
{\small Departamento de F\'{\i}sica Te\'orica, Facultad de Ciencias,
Universidad de Zaragoza,}\\ {\small Pedro Cerbuna 12, 50009, Zaragoza,
Spain.}}

\date{\today}

\maketitle

\begin{abstract}

In the study of the conformational behavior of complex systems, such
as proteins, several related statistical measures are commonly used to
compare two different potential energy functions. Among them, the
Pearson's correlation coefficient $r$ has no units and allows only
semi-quantitative statements to be made. Those that do have units of
energy and whose value may be compared to a physically relevant scale,
such as the root mean square deviation (RMSD), the mean error of the
energies (ER), the standard deviation of the error (SDER) or the mean
absolute error (AER), overestimate the distance between
potentials. Moreover, their precise statistical meaning is far from
clear. In this article, a new measure of the distance between
potential energy functions is defined which overcomes the
aforementioned difficulties. In addition, its precise physical meaning
is discussed, the important issue of its additivity is investigated
and some possible applications are proposed. Finally, two of these
applications are illustrated with
practical examples: the study of the van der Waals energy, as
implemented in CHARMM, in the \mbox{Trp-Cage} protein (PDB code 1L2Y)
and the comparison of different levels of the theory in the
ab initio study of the Ramachandran map of the model peptide
HCO-L-Ala-NH$_{2}$.
\vspace{0.2cm}\\ {\bf PACS:} 87.14.Ee, 87.15.-v, 87.15.Aa, 87.15.Cc,
89.75.-k
\vspace{0.2cm}\\

\end{abstract}

\section{Introduction}
\label{sec:introduction}

The most fundamental way to account for the behavior of a physical
system is through its energy function $H({\vec q},{\vec p})$, which
depends on the coordinates $\vec q$ and the momenta $\vec p$ of all
the particles. In normal situations, this function can be expressed as
the sum of the kinetic energy $K({\vec q},{\vec p})$\footnote{The
kinetic energy $K$ depends, in general, on the positions and the
momenta. However, if Cartesian coordinates are used, the dependence on
positions vanishes.} and the potential energy $V({\vec q})$. Because the
former is of general form for any type of system and, normally, it
does not affect the equilibrium properties, the latter is enough for a
complete characterization of the problem.
 
In the study and simulation of complex systems, such as proteins,
researchers often face the dilemma of choosing among many different
ways of calculating a conceptually unique potential energy $V$
\cite{PE:Fei2004JCC,PE:Lev2003JACS,PE:Nym2003PNAS,PE:Per2003JCC,PE:Onu2002JCC,PE:Dav2000JCC,PE:Edi1997JPCB}. Others
tackle the problem of designing new algorithms to perform this
calculation looking for the improvement of the relation between
accuracy and numerical complexity
\cite{PE:Gal2004JCC,PE:Pok2004PSC,PE:Bor2003JCP,PE:Im2003JCC,PE:Onu2000JPCB,PE:Dom1999JPCB,PE:Wag1999JCC,PE:Don1998JCC,PE:Gho1998JPCB,PE:Sca1998JPCB,PE:Qiu1997JPCA,PE:Sca1997JPCA}.

The energy $V$ studied may be the total potential energy of the system
or any of the terms in which it is traditionally
factorized\footnote{\label{foot:factorization}For example, in the case
of proteins \cite{PE:Alo2004BOOK,PE:Dil1999PSC}, some of the terms in
which the total potential energy is typically factorized are the
hydrogen-bonds energy, the van der Waals interaction, the excluded
volume repulsion, the Coulomb energy and the solvation energy.} and
the different ways of calculating it may stem from distinct
origins, namely, that different algorithms or approximations $A$ are
used, that the potential energy function depends on a number of free
parameters ${\vec P}$ or that it is computed on different but somehow
related systems $S$ (for a proper definition of this, see
Sec.~\ref{sec:pos_applications}).

Changes in these inputs produce different {\it instances} of the same
physical potential energy, which we denote by subscripting $V$. For
example, if it is calculated on the same system $S$, the algorithm and
approximations $A$ are held constant but two different set of
parameters ${\vec P}_{1}$ and ${\vec P}_{2}$ are used, our notation
made explicit would read as in the following
equations\footnote{\label{foot:analogous_def}Analogous definitions
may be made if different algorithms or approximations, $A_{1}$ and
$A_{2}$, are used or if $V$ is computed on two related systems, $S_{1}$
and $S_{2}$.}:

\begin{equation}
\label{eq:V1_V2}
V_{1}({\vec q}) := V(A,{\vec P}_{1},{\vec q},S)
\qquad \mathrm{and} \qquad
V_{2}({\vec q}) := V(A,{\vec P}_{2},{\vec q},S) \  .
\end{equation}

For each practical application of these two potential energy
functions, there is a limit on how different $V_{1}$ could be
from $V_{2}$ to preserve the relevant features of the system
under scrutiny. Clearly, if $V_{1}$ is {\it too distant} from $V_{2}$,
the key characteristics of the system behavior will be lost when
going from one function to the other.

In the literature, a number of different methods are used to quantify
this distance. Among them, the Pearson's correlation coefficient $r$
does not have units and its meaning is only semi-quantitative.  Some
others, such as the root mean square deviation (RMSD), the mean error of the
energies (ER), the standard deviation of the error (SDER) or the mean
absolute error (AER), do have units of energy and their values can be
compared to the physically relevant scale in each problem. However,
they tend to overestimate the sought distance even in the interesting
situations in which the potential energy functions under study are
physically proximate. The aim of this article is to define, justify
and describe a new meaningful measure $d(V_{1},V_{2})$ of the distance
between two instances of the same potential energy that overcomes the
aforementioned difficulties, and that allows to make precise
statistical statements about the way in which the energy differences
change when going from $V_{1}$ to
$V_{2}$\footnote{\label{foot:old_paper}The convenience of this
approach has been remarked in \cite{PE:Alo2004BPCP}. Note, however,
that in this article a different distance is defined.}.

In Sec.~\ref{sec:hypothesis}, the hypothesis made on $V_{1}$ and
$V_{2}$ to accomplish this are outlined and, in
Sec.~\ref{sec:definition}, the central object, $d(V_{1},V_{2})$, is
defined.  The statistical meaning of the distance herein introduced is
discussed in Sec.~\ref{sec:meaning} and \ref{sec:values} and some of
its possible applications to practical situations are proposed in
Sec.~\ref{sec:pos_applications}. A comparison to other commonly used
criteria is made and illustrated with a numerical example in
Sec.~\ref{sec:relation}. The important issue of the additivity of
$d(V_{1},V_{2})$, when the potentials studied are only a part of the
total energy, is investigated in Sec.~\ref{sec:additivity} and, in
Appendix~A, the metric properties of our distance are discussed. The
robustness of the van der Waals potential energy (as implemented in
CHARMM \cite{PE:Mac1998BOOK,PE:Bro1983JCC}) under a change in the
free parameters and the ab initio Ramachandran plots of 
HCO-L-Ala-NH$_{2}$ at different levels of the theory are studied in
Sec.~\ref{sec:application} as examples
of applications of the distance. Finally, Sec.~\ref{sec:conclusions}
is devoted to the conclusions and to a useful summary of the steps
that must be followed to use the distance in a practical case.

\section{Hypothesis}
\label{sec:hypothesis}

\begin{figure}[!ht]
\begin{center}
\epsfig{file=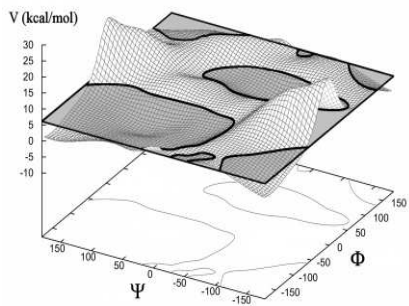,width=5.5cm}
\end{center}
\caption{\label{fig:v_const}{\small Space of constant
potential energy $V$ in a simple system with only two degrees of
freedom: an alanine dipeptide in vacuo. Potential energy surface (PES)
calculated ab initio at the \mbox{B3LYP/6-311++G(d,p)} level of the
theory in (Perczel, A. et al. 2003, {\it J. Comp. Chem.} 24:1026--1042).}}
\end{figure}

\begin{figure}[!ht]
\begin{center}
\epsfig{file=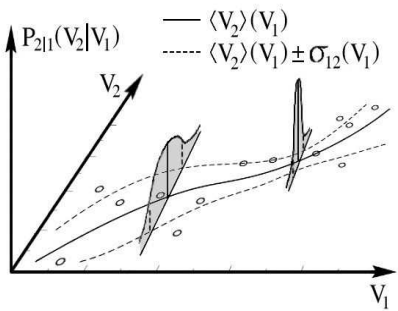,width=5.5cm}
\end{center}
\caption{\label{fig:gaussians}{\small Illustration of the
conditional-probability density function $P_{2|1}(V_{2} | V_{1})$ at
different $V_{1}$-positions. The points represent single
conformations of the system. The $V_{1}$-conditioned mean of $V_{2}$,
$\langle V_{2} \rangle (V_{1})$, is depicted as a solid line
and, although in the hypothesis it is assumed to depend linearly on
$V_{1}$, here it is shown as a more general function to
better illustrate the concepts involved. Analogously, the
$V_{1}$-conditioned standard deviation of $V_{2}$,
${\sigma}_{12}(V_{1})$ (which is assumed to be constant in the
hypothesis) is added to $\langle V_{2} \rangle (V_{1})$ (and
subtracted from it) and the result is depicted as broken
lines.}}
\end{figure}

In some cases traditionally studied in physics, the dependence of the
potential energy $V$ on the parameters is simple enough to allow a
closed functional dependence $V_{2}(V_{1})$ to be
found\footnote{\label{foot:simple_systems}For example, if the recovering
force constant of 
a harmonic oscillator is changed from $k_{1}$ to $k_{2}$, the
potential energy functions satisfy the linear relation
\mbox{$V_{2}({\vec q})=(k_{1}/k_{2})V_{1}({\vec q})$} for all the
conformations of the system; if the the atomic charges are rescaled by
a factor $\alpha$ (being actually ${\alpha}Q_{i}$) and $\alpha$ is
changed from ${\alpha}_{1}$ to ${\alpha}_{2}$, the free energies of
solvation calculated via the Poisson equation satisfy the linear
relation \mbox{$V_{2}({\vec
q})=({\alpha}_{1}/{\alpha}_{2})^{2}V_{1}({\vec q})$}, etc.}. However,
in the study of complex systems, such as proteins, this dependence is
often much more complicated, due to the high dimensionality of the
conformational space and to the fact that the energy landscape lacks
any evident symmetry. The set $C(V_{1})$ of the conformations with a
particular value of the potential energy $V_{1}$ typically spans large
regions of the phase space containing structurally different
conformations (see Fig.~\ref{fig:v_const}). When the system
is slightly modified, from $S_{1}$ to $S_{2}$, or an approximation is
performed (or the algorithm is changed), from $A_{1}$ to $A_{2}$, or
the free parameters are shifted, from ${\vec P}_{1}$ to ${\vec
P}_{2}$, each conformation ${\vec q}$ in $C(V_{1})$ is affected in a
different way and its potential energy is modified, from $V_{1}({\vec
q})$ to $V_{2}({\vec q})$, in a manner that does not depend trivially
on the particular region of the phase space which the conformation
${\vec q}$ belongs to. In such a case, a simple functional relation
$V_{2}(V_{1})$ is no longer possible to be found: for each value of
$V_{1}$, there corresponds now a whole distribution of values of
$V_{2}$ associated with conformations which share the same value of
$V_{1}$ but which are far apart in the conformational space.
Moreover, the projection of this \mbox{high-dimensional} \mbox{${\vec
q}$-space} into the \mbox{1-dimensional} \mbox{$V_{1}$-space} makes
it possible to treat $V_{2}$ as a random variable parametrically
dependent on $V_{1}$ (see Fig.~\ref{fig:gaussians}), in the
already suggested sense that, if one chooses at random a particular
conformation ${\vec q}_{i} \in C(V_{1})$, the {\it outcome} of the
quantity $V_{2}({\vec q}_{i})$ is basically
unpredictable\footnote{\label{foot:change_V1_by_V2}The same may be
said in the case that the conformations belong to $C(V_{2})$ and the
random variable is, in turn, $V_{1}$. The role of the two instances of
$V$ is interchangeable in the whole following reasoning, however, for
the sake of clarity, this fact will be made explicit in some cases and
will be tacitly assumed in others.}.

In this context, the hypothesis to be done about the two instances of
$V$ are, first, that the pair of values \mbox{$(V_{1}({\vec
q}_{i}),V_{2}({\vec q}_{i}))$} is independent of \mbox{$(V_{1}({\vec
q}_{j}),V_{2}({\vec q}_{j}))$} if $i \ne j$ and, second, that
the probability distribution of $V_{2}$ conditioned by $V_{1}$ is
normal with mean $b_{12}V_{1}+a_{12}$ and standard deviation
${\sigma}_{12}$ and that, conversely, the probability distribution of
$V_{1}$ conditioned by $V_{2}$ (i.e., the distribution of the random
variable $V_{1}$ in the space $C(V_{2})$, analogous to $C(V_{1})$) is
normal with mean $b_{21}V_{2}+a_{21}$ and standard deviation
${\sigma}_{21}$. Where $a_{12}$, $b_{12}$, ${\sigma}_{12}$, $a_{21}$,
$b_{21}$ and ${\sigma}_{21}$ are constants not dependent on $V_{1}$ or
$V_{2}$. This can be summarized in the following expressions for the
conditional-probability density functions:

\begin{subequations}
\label{eq:cond_pdfs_V1V2}
\begin{align}
& P_{2|1}(V_{2} | V_{1})=\frac{1}{\sqrt{2\pi}{\sigma}_{12}}
    \exp \left [ -\frac{(V_{2}-(b_{12}V_{1}+a_{12}))^{2}}
	 {2{\sigma}_{12}^{2}} \right ] \  , \label{eq:cond_pdfs_V1V2_a}\\
& P_{1|2}(V_{1} | V_{2})=\frac{1}{\sqrt{2\pi}{\sigma}_{21}}
    \exp \left [ -\frac{(V_{1}-(b_{21}V_{2}+a_{21}))^{2}}
	 {2{\sigma}_{21}^{2}} \right ] \  . \label{eq:cond_pdfs_V1V2_b}
\end{align}
\end{subequations}

In general, one may reason about the whole conformational space of the
system $C$ and regard each randomly selected conformation ${\vec
q}_{i}$ as a single {\it numerical experiment} to which the value of
two random variables, $V_{1}({\vec q}_{i})$ and $V_{2}({\vec q}_{i})$,
can be assigned. However, no hypothesis need to be made about the
joint probability density function
$P_{12}(V_{1},V_{2})$\footnote{\label{foot:binormal}The hypothesis
that $P_{12}(V_{1},V_{2})$ is bivariate normal, for example, is
stronger than the assumptions in Eq.~\ref{eq:cond_pdfs_V1V2}.
The latter can be derived from the former.}. For the distance herein
introduced to be meaningful, it suffices to assume
Eq.~\ref{eq:cond_pdfs_V1V2}.

Regarding the question of whether in a typical case these hypothesis
are fulfilled or not, some remarks should be made. First, the
satisfaction of the independence hypothesis depends mainly on the
process through which the working set of conformations $\{{\vec
q}_{i}\}^{N}_{i=1}$ is generated. For example, if the conformations
are extracted from a single molecular dynamics trajectory letting only
a short simulation time pass between any pair of them, their energies
will be obviously correlated and the independence will be broken. If,
on the contrary, each conformation ${\vec q}_{i}$ is taken from a
different trajectory (see the first example in
Sec.~\ref{sec:application}), one may reasonably expect this assumption
to be fulfilled, i.e., the independence hypothesis is normally under
researcher's control.

\begin{figure}[!ht]
\begin{center}
\epsfig{file=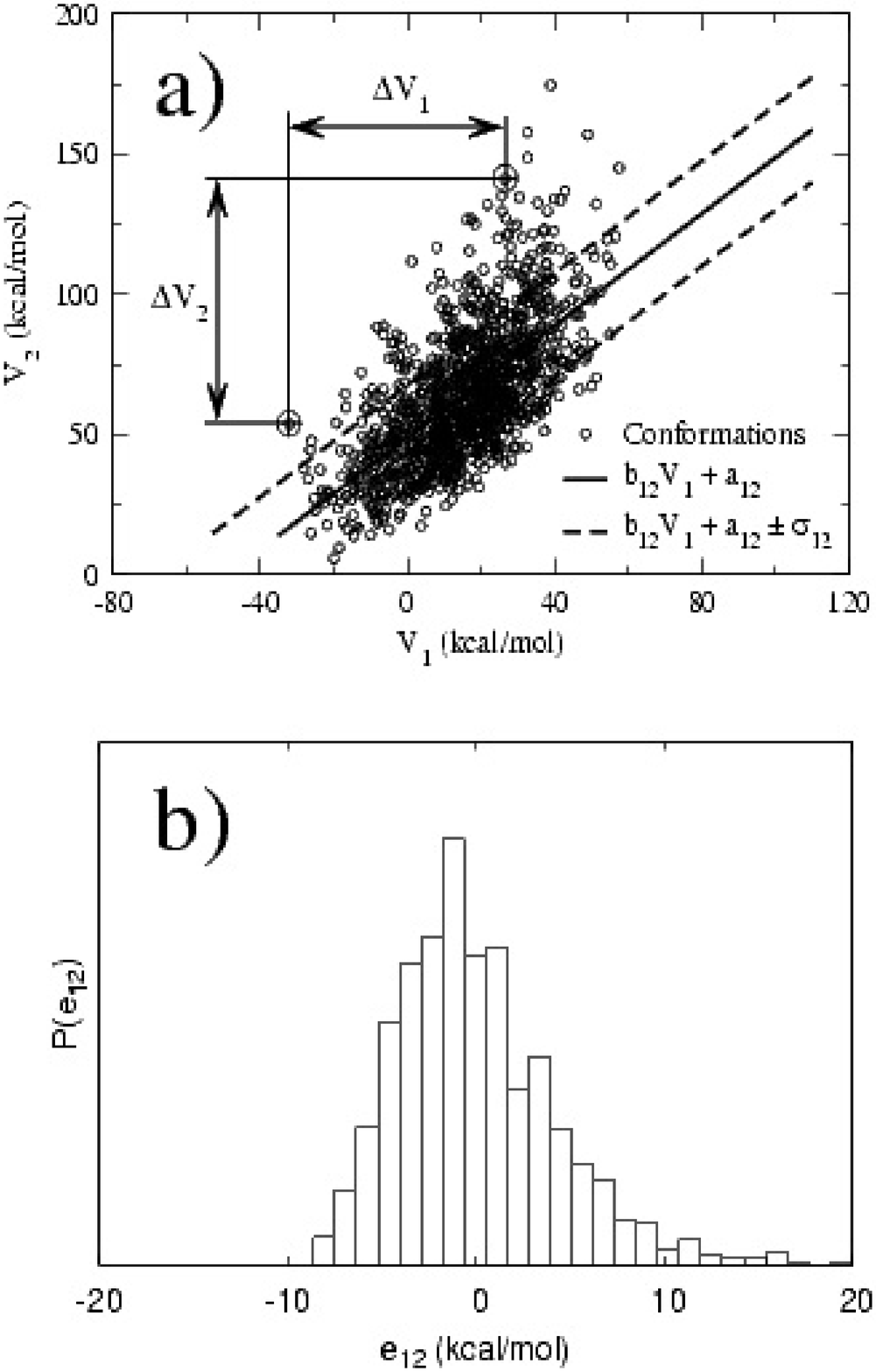,width=5.5cm}
\end{center}
\caption{\label{fig:hyp_exp}{\small {\bf (a)} $V_{1}$- and
\mbox{$V_{2}$-energies} of the set of $1100$ conformations of the
\mbox{Trp-Cage} protein used in the first example of
Sec.~\ref{sec:application}. Both potentials are the van der Waals
energy as implemented in CHARMM; $V_{1}$ corresponds to
\mbox{$R_{C}=1.275$ \AA} and $V_{2}$ to \mbox{$R_{C}=3.275$ \AA},
${\varepsilon}_{C}$ is fixed to $-0.020$ kcal/mol. The values of
${\Delta}V_{1}$ and ${\Delta}V_{2}$ for a selected pair of
conformations are also depicted. The solid line represents the
least-squares fit and the region where the probability of finding a
conformation is largest is enclosed by broken lines. {\bf (b)}
Histogram of the residues \mbox{$e_{12}({\vec q}_{i}):=V_{2}({\vec
q}_{i})- [b_{12}V_{1}({\vec q}_{i})+a_{12}]$} associated to the
mean-squares fit in Fig. {\it a}.}}
\end{figure}

The normality hypothesis, however, is of a different nature.  That the
distribution of $V_{2}$ be normal for a particular value of $V_{1}$
may be thought as a consequence of the large number of degrees of
freedom the system possesses, of the usual pairwise additivity of the
forces involved and of the Central Limit Theorem (this, in fact, can
be proved in some simple cases). Nevertheless, that
the $V_{1}$-conditioned mean of $V_{2}$, $\langle V_{2} \rangle
(V_{1})$, is linear in $V_{1}$ and that the $V_{1}$-conditioned
standard deviation of $V_{2}$, ${\sigma}_{12}(V_{1})$, is a constant
must be regarded as a zeroth order approximation that should be
checked in each particular case (see Fig.~\ref{fig:gaussians}).
It is worth pointing out, however, that, for the commonly used
statistical quantities $r$, RMSD, etc. to be useful, this
assumption must also be made (see Sec.~\ref{sec:relation})
and also that it has been found to be approximately fulfilled in
several cases studied (see, for example, \cite{PE:Alo2004BPCP}
and Fig.~\ref{fig:hyp_exp}{\it b}).

\section{Definition}
\label{sec:definition}

For the aforementioned cases in which the dependence of the potential
energy on the parameters is simple enough (see
footnote~\ref{foot:simple_systems}), one can describe $V_{2}(V_{1})$
by a closed analytical formula and exactly compute $a_{12}$, $b_{12}$
and ${\sigma}_{12}$ (this last quantity being equal to zero in such a
situation). However, in a general situation, the parameters
entering Eq.~\ref{eq:cond_pdfs_V1V2} can not be calculated
analytically. In such a case, one may at most have a finite collection
of $N$ conformations $\{{\vec q}_{i}\}^{N}_{i=1}$ and the respective
values $V_{1}({\vec q}_{i})$ and $V_{2}({\vec q}_{i})$ for each one
of them.

From this finite knowledge about the system, one may statistically
estimate the values of $a_{12}$, $b_{12}$ and ${\sigma}_{12}$. Under
the hypothesis assumed in the previous section, the least-squares
estimators \cite{PE:Bev2003BOOK,PE:Pre2002BOOK} of these quantities
are optimal in the precise statistical sense that they are
maximum-likelihood and have minimum variance in the class of linear
and unbiased estimators\footnote{\label{foot:estimators}The same
letters are used for the ideal parameters $a_{12}$, $b_{12}$, and
${\sigma}_{12}$ and for their least-squares best estimators, because the
only knowledge that one may have about the former comes from the
calculation of the latter.} \cite{PE:Lie1967BOOK}.

If we denote $V_{1}^{i}:=V_{1}({\vec q}_{i})$ and
$V_{2}^{i}:=V_{2}({\vec q}_{i})$, and $N$ is the number of
conformations in the working set, the mean-squares maximum-likelihood
estimators\footnote{\label{foot:sig_estimator}In this article, the
maximum-likelihood estimator for ${\sigma}_{12}$ (with $N$ in the
denominator) is preferred to the unbiased one (with $N-2$ in the
denominator) for consistency. Anyway, for the values of $N$ typically
used, the difference between them is negligible.}
of $a_{12}$, $b_{12}$ and ${\sigma}_{12}$ are given by the
following expressions \cite{PE:Bev2003BOOK,PE:Pre2002BOOK}:

\begin{subequations}
\label{eq:estimators}
\begin{align}
& b_{12}=\frac{\mathrm{Cov}(V_{1},V_{2})}{{\sigma}_{1}^{2}} \  ,
\label{eq:estimators_a} \\
& a_{12}={\mu}_{2}-b_{12}{\mu}_{1} \  , \label{eq:estimators_b} \\
& {\sigma}_{12}=\left [ \frac{{\sum}_{i=1}^{N}
    (V_{2}^{i}-(b_{12}V_{1}^{i}+a_{12}))^{2}}{N} \right ]^{1/2} \  ,
\label{eq:estimators_c}
\end{align}
\end{subequations}

where:

\begin{subequations}
\label{eq:stat_quantities}
\begin{align}
{\mu}_{1}&:=\frac{1}{N}\sum_{i=1}^{N}V_{1}^{i} \  , 
\label{eq:stat_quantities_a} \\
{\mu}_{2}&:=\frac{1}{N}\sum_{i=1}^{N}V_{2}^{i} \  , 
\label{eq:stat_quantities_b} \\
{\sigma}_{1}&:=\left [ \frac{1}{N}\sum_{i=1}^{N}(V_{1}^{i}-{\mu}_{1})^{2}
  \right ]^{1/2} \  ,
\label{eq:stat_quantities_c} \\
\mathrm{Cov}(V_{1},V_{2})&:=\frac{1}{N}\sum_{i=1}^{N}
  (V_{1}^{i}-{\mu}_{1})(V_{2}^{i}-{\mu}_{2}) \  .
\label{eq:stat_quantities_d}
\end{align}
\end{subequations}

The quantities with $21$ subscripts are found by changing \mbox{$1
\leftrightarrow 2$} in the preceding expressions and the central
object of this article, the distance $d(V_{1},V_{2})$ between two
different instances of the same potential energy $V$ is defined as
follows:

\begin{equation}
\label{eq:def_d}
d(V_{1},V_{2}) := \left ( {\sigma}_{12}^{2}+
  {\sigma}_{21}^{2} \right )^{1/2} \  .
\end{equation}

It must be stressed here that the measured distance depends on the
working set, $\{{\vec q}_{i}\}^{N}_{i=1}$, of conformations
chosen. More precisely, it depends on the occurrence probability of an
arbitrary conformation $\vec{q}$ in the set. This probability must be
decided a priori from considerations regarding which regions of the
phase space are more relevant to answer the questions posed and up to
what extent. For example, if one believes the system under study to
be in thermodynamical equilibrium, then, it would be reasonable to
generate a working set in which the probability that $\vec{q}$ occurs
is proportional to its Boltzmann weight. If, on the contrary, one
doubts whether the system is ergodic or not (as in the case of
proteins) or one simply wants to study in detail the dynamical
trajectories out of equilibrium, then, all the conformations
in the phase space should be weighted equally and the probability
should be flat.

\section{Meaning}
\label{sec:meaning}

Under the hypothesis made in Sec.~\ref{sec:hypothesis}
(independence and Eq.~\ref{eq:cond_pdfs_V1V2}), a simple expression
may be written for the probability density function of the
$V_{2}$-energy differences ${\Delta}V_{2}$ conditioned by the
knowledge of the $V_{1}$-energy differences ${\Delta}V_{1}$:

\begin{equation}
\label{eq:cond_pdf_DV2}
P_{{\Delta}2|{\Delta}1}({\Delta}V_{2} | {\Delta}V_{1}) =
\frac{1}{\sqrt{2\pi}d_{12}}
    \exp \left [ -\frac{({\Delta}V_{2}-b_{12}{\Delta}V_{1})^{2}}
	 {2d_{12}^{2}} \right ] \  .
\end{equation}

The quantity $d_{12}$ in this equation is defined as follows:

\begin{equation}
\label{eq:def_d12}
d_{12}:=\sqrt{2}{\sigma}_{12} \  .
\end{equation}

It is related to the distance defined in Eq.~\ref{eq:def_d} via the
following expression:

\begin{equation}
\label{eq:d12}
d(V_{1},V_{2}) = \left ( \frac{d_{12}^{2}+d_{21}^{2}}{2} \right )^{1/2} \  .
\end{equation}

And it encodes the {\it loss of information} involved in the transit
from $V_{1}$ to $V_{2}$ through the following important
properties:

\begin{enumerate}
\item The addition of an energy reference shift $a_{12}$ between
$V_{1}$ and $V_{2}$ has neither an implication in the physical
behavior of the system nor in the numerical value of $d_{12}$.

\item One of the novel features of the distance herein defined is that
no loss of information is considered to occur (i.e., $d_{12}=0$) if
there is only a constant rescaling $b_{12}$ between the two potential
energy functions studied. Although such a transformation does have
physical implications and would change the transition rates in a
molecular dynamics simulation and alter the effective temperature in
any typical Monte Carlo algorithm, $V_{1}$ can be easily recovered
from $V_{2}$, if pertinent, upon division of $V_{2}$ by $b_{12}$. If
the two potential energy functions are on equal footing (e.g., they
correspond to different values of the free parameters (see
Sec.~\ref{sec:pos_applications})), there is no {\it correct} energy
scale defined. However, in the case that the distance is used to
compare an approximated potential to a more ab initio one or even to
experimental data, the {\it correct} energy scale must be considered
to be that of the more reliable potential and the rescaling $b_{12}$
may be safely removed as indicated above. Note that the quantity
$d_{12}$ changes, when this removal is performed, from $d_{12}$ to
$d_{12}/|b_{12}|$ (to see this, take the analogous for $\sigma_{2}$ of
Eq.~\ref{eq:stat_quantities_c} and change $V_{2}^{i}$ by
$V_{2}^{i}/b_{12}$, finally, take the result to
Eq.~\ref{eq:d12_vs_r12_a}) and it is the second value which must be
considered as the relevant one.

\item \label{point:linear} Directly from its very definition in
Eq.~\ref{eq:def_d12}, one has that $d_{12}=0$ is equivalent to $V_{2}$
being exactly a linear transformation of $V_{1}$, i.e., to
$V_{2}({\vec q}_{i})=b_{12}V_{1}({\vec q}_{i})+a_{12}$, \mbox{$\forall
{\vec q}_{i} \in \{{\vec q}_{i}\}^{N}_{i=1}$}.

\item \label{point:meaning} As stated above, $b_{12} \ne 1$ and/or
$a_{12} \ne 0$ must be regarded as two different types of systematic
error easily removable and not involving any loss of information when
one changes $V_{1}$ by $V_{2}$. In the general case, however, the
energy differences associated to each potential energy function (which
are the relevant physical quantities that govern the system behavior)
present an additional random error which is intrinsic to the
discrepancies between the potentials and can not be removed. As can be
seen in Eq.~\ref{eq:cond_pdf_DV2}, in this situation, $d_{12}$ is the
standard deviation of the random variable ${\Delta}V_{2}$ and, as its
value decreases, the distribution becomes sharper around the average
$b_{12}{\Delta}V_{1}$. Moreover, because the distribution is normal,
the probability of ${\Delta}V_{2}$ being in the interval
\mbox{$(b_{12}{\Delta}V_{1}-K d_{12},b_{12}{\Delta}V_{1}+K d_{12})$}
is $\sim 38\%$ for $K=1/2$, $\sim 68\%$ for $K=1$, $\sim 95\%$ for
$K=2$, etc. Hence, $d_{12}$ quantifies the random error between the
trivially transformed potential $b_{12}V_{1}+a_{12}$ and $V_{2}$,
i.e., the unavoidable and fundamentally statistical part of the
difference between $V_{1}$ and $V_{2}$ which stems from the complex
character of the system.

\item To gain some insight about the meaning of $d_{12}$, the
following {\it gedanken experiment} may be performed: if a Gaussian
{\it noise} with zero mean and variance equal to $s^{2}$ were
independently added to the linearly transformed $V_{1}$-energy,
\mbox{$b_{12}V_{1}({\vec q})+a_{12}$}, of each conformation and the
resulting potential were denoted by $V_{2}$, then one would have that
the hypothesis in Eq.~\ref{eq:cond_pdfs_V1V2_a} is fulfilled and that
$d_{12}=\sqrt{2}s$. Therefore, $d_{12}$ may be regarded (except for a
harmless factor $\sqrt{2}$) as the size of the Gaussian noise arising
in the whole energy landscape when one changes $b_{12}V_{1}+a_{12}$ by
$V_{2}$.

\item Closely related to the properties in the two preceding points,
an illuminating statistical statement about the energetic
ordering of the conformations can be derived from
Eq.~\ref{eq:cond_pdf_DV2}. The probability that the energetic order
of two randomly selected conformations is maintained when going
from $V_{1}$ to $V_{2}$ (more precisely, that
\mbox{$sign({\Delta}V_{2})=sign(b_{12}{\Delta}V_{1})$}) conditioned by
the knowledge of ${\Delta}V_{1}$, can be easily shown to be:

\begin{equation}
\label{eq:P_ord}
P_{\mathrm{ord}} \left ( \frac{|b_{12}{\Delta}V_{1}|}{d_{12}} \right ) =
    \frac{1}{2}+
    \frac{1}{\sqrt{2\pi}} \int_{0}^{|b_{12}{\Delta}V_{1}|/d_{12}}
    \exp \left [ -\frac{x^{2}}{2} \right ] \mathrm{d}x \  .
\end{equation}

The intuitive meaning of this expression is that $d_{12}$ is the
$V_{1}$-energy difference at which two randomly selected conformations
can be typically {\it resolved} using $V_{2}$ after the removal of the
harmless rescaling $b_{12}$ (see Pt.~\ref{point:meaning}). Certainly,
if one has that \mbox{$|b_{12}{\Delta}V_{1}| \ll d_{12}$}, then
\mbox{$P_{\mathrm{ord}}=1/2$}, reflecting a total lack of knowledge
about the sign of ${\Delta}V_{2}$ and, consequently, $V_{2}$ can not
be used to resolve $V_{1}$-energy differences. If, on the contrary,
\mbox{$|b_{12}{\Delta}V_{1}| \gg d_{12}$}, then
\mbox{$P_{\mathrm{ord}}=1$} and the conformations ordering is exactly
conserved. In any intermediate point, $P_{\mathrm{ord}}$ is a rapidly
increasing function of \mbox{$|b_{12}{\Delta}V_{1}|/d_{12}$} that
reaches a {\it reasonable} value ($\sim 84\%$) when its argument
equals $1$, i.e., when \mbox{$|b_{12}{\Delta}V_{1}|=d_{12}$}. Some
other interesting points are \mbox{$P_{\mathrm{ord}}(1/2) \simeq
69\%$} or \mbox{$P_{\mathrm{ord}}(2) \simeq 98\%$}.

\item Finally, some clarifying properties of the distance associated
to its relation to the Pearson's correlation coefficient will be
investigated in Sec.~\ref{sec:relation} (see specially
Eq.~\ref{eq:d12_vs_r12_a}).
\end{enumerate}

The same considerations may be done about $d_{21}$ regarding the
transit from $V_{2}$ to $V_{1}$ and, as can be seen in
Eq.~\ref{eq:d12}, the square of $d(V_{1},V_{2})$ is the mean of the
squares of $d_{12}$ and $d_{21}$. Therefore, this measure of the
difference between potential energy functions quantifies the average
size of the uncertainty in the energy differences of the system that
arises from changing one of the potentials studied by the other. If
the comparison is performed between potential energy functions that
stand on the same footing (see, for example, the second possible
application in Sec.~\ref{sec:pos_applications}), the symmetric
quantity $d(V_{1},V_{2})$ should be used as a summarizing measure of
the loss of information involved in the transit from $V_{1}$ to
$V_{2}$ and vice versa. However, if one of the potentials is a priori
considered to be more ab initio or more accurate and it is compared to
a less reliable instance, $V_{1}$ may denote the former, $V_{2}$ the
latter and one may use only $d_{12}$ as the measure
of the discrepancies between
them\footnote{\label{foot:d12_eq_d21}Note, from Eq.~\ref{eq:d12},
that, if $d_{12}=d_{21}$, then
\mbox{$d(V_{1},V_{2})=d_{12}=d_{21}$}.}.

Hence, although both the
discussion regarding the relevant values of the distance in the
following section and the investigation of its mathematical properties
in Sec.~\ref{sec:additivity} and Appendix~A are referred to
$d(V_{1},V_{2})$ for generality, they may be equally applied to
$d_{12}$. Conversely, the comparison between $d_{12}$ and the quantities
commonly used in the literature done in Sec.~\ref{sec:relation} may be
extended to $d(V_{1},V_{2})$ upon symmetrization of ER, which
is the only asymmetrical one.

\section{Relevant values of the distance}
\label{sec:values}

Regarding the value of $d(V_{1},V_{2})$ in a practical case, some
remarks must be made. One may expect two special values of the
distance to exist: $d_{\mathrm{min}}$ and $d_{\mathrm{max}}$. In such
a way that, if \mbox{$d(V_{1},V_{2})<d_{\mathrm{min}}$}, one potential
energy function may be substituted by the other without altering the
key characteristics of the system behavior, and that, if
\mbox{$d(V_{1},V_{2})>d_{\mathrm{max}}$}, then, the substitution is
not acceptable.  This limiting values must be set depending on the
particularities of the system studied and on the questions sought to
be answered, and it may even be the case that some special features of
the energy landscape are the main responsible of the behavior under
scrutiny. For example, we are not going to establish any strict limit
on the accuracy required for a potential energy function to
successfully predict the folding of proteins
\cite{PE:Alo2004BOOK,PE:Dil1999PSC}. We consider this question a
difficult theoretical issue, whose solution probably requires a much
deeper knowledge of the protein folding problem itself than the one
that exists at present, and we believe that it may be possible a
priori that some special features of the energy landscapes of proteins
(such as a \mbox{funnel-like} shape
\cite{PE:Onu2000APC,PE:Dob1998ACIE}) are the main responsible of the
high efficiency and cooperativity of the folding process
\cite{PE:Alo2004BOOK,PE:Dil1999PSC}. If this were the case, a
different procedure for measuring the distance between potential
energy functions could be devised for this situation
\cite{PE:Per1997FD,PE:Per1996FD,PE:Pan1995JCP}, as any change of
$V_{1}$ by $V_{2}$ which did not significantly alter these special
features would be valid even if the value of $d(V_{1},V_{2})$ were
very large. Our definition of $d(V_{1},V_{2})$, being based in
characteristics shared by many complex systems and statistically
referred to the whole energy landscape, is of more general application
but cannot detect such particular features as the ones mentioned.

However, due to the laws of Statistical Mechanics, a rather stringent
but general value for $d_{\mathrm{min}}$ can be used to a priori
assess the interchangeability of $V_{1}$ and $V_{2}$.  As can be seen
in the thermodynamical equilibrium Boltzmann distribution, in which
the probability $p_{i}$ of a conformation ${\vec q}_{i}$ is
proportional to $\exp (-V({\vec q}_{i})/RT)$, the order of the
physical uncertainty in the potential energies of a system in contact
with a thermal reservoir at temperature $T$ is given by the quantity
$RT$\footnote{\label{foot:RT}$RT$ is preferred to $k_{B}T$ because per
mole energy units are used in this article.}.  This typical energy
sets the scale of the thermal fluctuations and it also determines the
transition probability, \mbox{$\mathrm{min}[1,\exp (-(V({\vec
q}_{i+1})-V({\vec q}_{i}))/RT)]$}, in the Metropolis Monte Carlo
scheme and the spread of the stochastic term in the Langevin equation
\cite{PE:Lea2001BOOK,PE:Rot1997BOOK}. Consequently, in this
article, $RT$ (which equals \mbox{$\sim 0.6$ kcal/mol} at room
temperature) will be used as a general lower bound for
$d_{\mathrm{min}}$.  The results will be presented in units of $RT$
and any two instances $V_{1}$ and $V_{2}$ of the same potential energy
function whose distance $d(V_{1},V_{2})$ be smaller than $RT$ will be
regarded as physically equivalent\footnote{\label{foot:chem_acc}This
discussion is closely related to the common use of the concept of {\it
chemical accuracy}, typically defined in the field of ab initio
quantum chemistry as predicting bond-breaking energies to \mbox{$1$
kcal/mol} \cite{PE:Day2003PRL}.}.

Regarding $d_{max}$, no estimations of its value can a priori be made
without referring to the particular potential energy functions
compared and the relevant behavior studied. The fact that
Eq.~\ref{eq:d12_vs_r12_b} has an absolute maximum when $r_{12}=0$
sets only the worst possible upper bound and is only of
mathematical interest.

\section{Possible applications}
\label{sec:pos_applications}

There are at least three basic situations in which the distance
defined in this article may be used to quantify the discrepancies
between two different instances, $V_{1}$ and $V_{2}$, of the same
potential energy:

\begin{itemize}
\item If the difference between $V_{1}$ and $V_{2}$ arise from the use
of two distinct algorithms or approximations, $d(V_{1},V_{2})$ (or
$d_{12}$, see the final lines of Sec.~\ref{sec:meaning}) may help us
to decide whether the less numerically complex instance could be used
or not. For example, one may compare the electrostatic part of the
solvation energy calculated via solving the Poisson equation
\cite{PE:Rou2001BOOK,PE:Oro2000CR,PE:Zha1997JFI,PE:Hon1995SCI} with
the instance of the same energy calculated using one of the many
implementations of the Generalized Born model
\cite{PE:Pok2004PSC,PE:Im2003JCC,PE:Bas2000ARPC,PE:Onu2000JPCB,PE:Dom1999JPCB,PE:Gho1998JPCB,PE:Sca1997JPCA,PE:Sti1990JACS},
which are much less computationally demanding and more suitable for
simulating macromolecules. If the distance between them is small
enough for the behavior under study not to be much modified (see
Sec.~\ref{sec:values}), the latter could be used. The second example
in Sec.~\ref{sec:application} is devoted to illustrate this type of
comparison, which, of the three possible applications proposed in
this section, is the one most commonly found in the literature.

\item If the algorithm and the approximations are
fixed and only one system $S$ is studied, any reasonable functional
form used to account for $V$ will be a simplified model of physical
reality and it will contain a number of free parameters ${\vec
P}$. These parameters, which, in most of the cases, are not physically
observable, must be fit against experimental or more ab initio
results before using the function for practical purposes.
For any fit to yield statistically significant values of the
parameters, the particular region of the parameter space in which the
final result lies must have the property of {\it robustness}, i.e., it
must occur that, if the found set of parameters values is slightly
changed, then, the relevant characteristics of the potential energy
function which depends on them are also approximately kept
unmodified. If this were not the case, a new fit, performed using a
different set of experimental (or more ab initio) points, could
produce a very distant potential. If $V_{1}$ and $V_{2}$ come from the
same family of potential energy functions and they correspond to
different values of the free parameters ${\vec P}$, the distance
$d(V_{1},V_{2})$ between them may help us to assess the robustness of
the potentials under changes on the parameters. In the first example of
Sec.~\ref{sec:application}, the robustness of the van der Waals
potential energy implemented in the well-known molecular dynamics
program CHARMM \cite{PE:Bro1983JCC,PE:Mac1998BOOK} is quantified as
an example of this.

\item The last application of the distance is fundamentally different
of the ones in the previous points but, although the reasoning
throughout the article is intentionally biased, for the sake of
clarity, toward the study of potential energies of the same system,
one may appeal to the same underlying assumptions to compare two
different systems, $S_{1}$ and $S_{2}$, provided that a meaningful
mapping can be established between both conformational
spaces\footnote{\label{foot:conf_xy}In short, for the distance
criterion to be applied, one needs to be able to assign two energies,
$V_{1}({\vec q})$ and $V_{2}({\vec q})$, to each conformation ${\vec
q}$. This is done trivially in the first two points but it requires a
mapping between the conformational spaces of $S_{1}$ and $S_{2}$ in
the third case.}. For example, if the conformations of a particular
protein are described only by their backbone angles, one can define an
unambiguous mapping between the conformations of, say, the wild-type
chain and any mutated form, in such a way that $V_{1}({\vec q})$ would
represent the energies of the former and $V_{2}({\vec q})$ those of
the latter. The distance $d(V_{1},V_{2})$, in this case, quantifies
how different the energy landscapes of the two systems are and,
depending on the features under study, how sensitive the behavior of
the protein is to mutations. The comparison of a potential, $V_{1}$, to
another one, $V_{2}$, which comes from the first via {\it
integrating-out} certain degrees of freedom and which is commonly
termed {\it effective potential energy} \cite{PE:Laz2003BPC}, may
be considered to be another example of this type of application.
\end{itemize}

\section{Relation to other statistical quantities}
\label{sec:relation}

In the literature, some comparisons between
potentials\footnote{\label{foot:comparisons}It must be pointed out
that we have only found in the literature examples of the comparison
between two potentials corresponding to the first case described in
Sec.~\ref{sec:pos_applications}, which is associated to different
algorithms or approximations. No examples of robustness studies have
been found and, regarding the third case, in which the differences
arise from a slight change in the system, such as a mutation in a
protein, only articles investigating the total free energy of folding
have been found \cite{PE:Cam2004JMB,PE:Lin2003BPJ}.} are performed a
posteriori, i.e., not directly studying the energies but computing
some derived quantities, such as the $pK_{a}$ of titratable groups
\cite{PE:Pok2004PSC,PE:Onu2000JPCB}, investigating molecular dynamics
trajectories
\cite{PE:Gal2004JCC,PE:Nym2003PNAS,PE:Dav2000JCC,PE:Dom1999JPCB},
comparing the ability of the different instances of $V$ to select the
correct native state of a protein from a set of decoys
\cite{PE:Gal2004JCC}, etc.

For the a priori comparison of two ways of calculating the same
potential energy, one may investigate the whole energy landscape
visually if the system has no more than two degrees of freedom
\cite{PE:Wag1999JCC,PE:Sca1998JPCB}, but, if the object of study is a
protein or another complex system, the vastness of the conformational
space and its lack of symmetries require the utilization of
statistical quantities calculated from the energies of a finite set of
conformations. Among the most common such measures, one may find the
root mean square deviation (RMSD)
\cite{PE:Fei2004JCC,PE:Dav2000JCC,PE:Dom1999JPCB,PE:Wag1999JCC,PE:Gho1998JPCB},
the mean error of the energies (ER)
\cite{PE:Fei2004JCC,PE:Bor2003JCP,PE:Edi1997JPCB}, the standard
deviation of the error (SDER) \cite{PE:Edi1997JPCB}, the mean of the
absolute error (AER) \cite{PE:Im2003JCC}, all of which have units of energy,
and the Pearson's correlation coefficient $r$
\cite{PE:Bor2003JCP,PE:Lev2003JACS,PE:Per2003JCC,PE:Dav2000JCC,PE:Don1998JCC,PE:Edi1997JPCB,PE:Sca1997JPCA},
which does not have units. Finally, in \cite{PE:Gho1998JPCB}, a root
mean square of the difference in the relative energies (REL) (see
Eq.~\ref{eq:quantities_e} for a clarification) which is proximate to
$d_{12}$ is defined, however, it has not been found to be used
in any other work.

If we use the same notation as in Eq.~\ref{eq:estimators} and
we define \mbox{${\Delta}V_{12}^{i}:=V_{2}^{i}-V_{1}^{i}$}, the
statistical quantities mentioned in the preceding paragraph (except
$r$, which will be discussed later) are given by the expressions:

\begin{subequations}
\label{eq:quantities}
\begin{align}
\mathrm{RMSD}(V_{1},V_{2})&:=\left [ \frac{1}{N}\sum_{i=1}^{N}
  ({\Delta}V_{12}^{i})^{2} \right ]^{1/2} \ , \label{eq:quantities_a} \\
\mathrm{ER}(V_{1},V_{2})&:=\frac{1}{N}\sum_{i=1}^{N}{\Delta}V_{12}^{i} \ , 
\label{eq:quantities_b} \\
\mathrm{SDER}(V_{1},V_{2})&:=\left [ \frac{1}{N}\sum_{i=1}^{N}
  ({\Delta}V_{12}^{i}-\mathrm{ER}(V_{1},V_{2}))^{2} \right ]^{1/2} \ ,
\label{eq:quantities_c} \\
\mathrm{AER}(V_{1},V_{2})&:=
  \frac{1}{N}\sum_{i=1}^{N}|{\Delta}V_{12}^{i}| \  , 
\label{eq:quantities_d} \\
\mathrm{REL}(V_{1},V_{2})&:=\left [ \frac{2}{N(N-1)}\sum_{i=1}^{N}
  \sum_{j=i+1}^{N}({\Delta}V_{12}^{j}-{\Delta}V_{12}^{i})^{2}
  \right ]^{1/2} \ .
\label{eq:quantities_e}
\end{align}
\end{subequations}

\begin{figure}[!ht]
\begin{center}
\epsfig{file=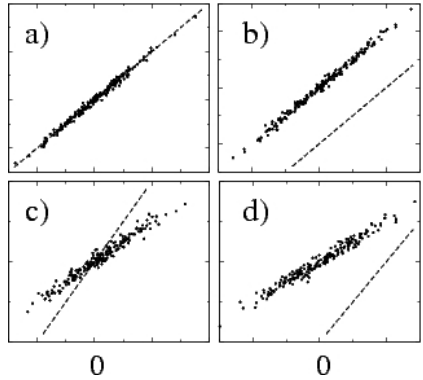,width=5.5cm}
\end{center}
\caption{\label{fig:numerical}{\small Numerical examples of the
possible situations found in practical problems. $200$ conformations
are depicted for each case with the values of $V_{1}$ in the $x$-axis
and the ones of $V_{2}$ in the $y$-axis (both in arbitrary energy
units). The broken line corresponds to the line $V_{2}=V_{1}$. {\bf
(a)} $b_{12} \simeq 1$ and $a_{12} \simeq 0$. {\bf (b)} $b_{12} \simeq
1$ and $a_{12} \simeq 200$. {\bf (c)} $b_{12} \simeq 1/2$ and $a_{12}
\simeq 0$. {\bf (d)} $b_{12} \simeq 1/2$ and $a_{12} \simeq 200$. }}
\end{figure}

In the following points, this measures of the difference between
potential energy functions are individually compared to the distance
defined in Sec.~\ref{sec:definition} and their limitations with
respect to $d_{12}$ are pointed out\footnote{\label{foot:diff_d12}The
quantity ER is not symmetrical. This is why all the measures in
Eq.~\ref{eq:quantities} are compared to $d_{12}$ and not to its
symmetrized version $d(V_{1},V_{2})$ (see Appendix~A and
the final part of Sec.~\ref{sec:meaning}).}:

\begin{itemize}
\item The first one, the RMSD, which is one of the most commonly used,
presents the major flaw of overestimating the importance of an energy
reference shift between $V_{1}$ and $V_{2}$. This transformation,
which has no physical implications in the conformational behavior of
the system, must not influence the assessment of the difference
between potentials. This fact is, for example, detected in some of the
comparisons performed in \cite{PE:Dav2000JCC} and recognized to be
conceptually erroneous in \cite{PE:Fei2004JCC}, where the shift is
removed by minimizing the RMSD with respect to it. In addition, the
RMSD also overestimates the effect of a slope $b_{12} \ne 1$ between the two
potentials, a fact that, as has been remarked in
Sec.~\ref{sec:meaning}, is not desirable (for a practical case in
which the loss of information is small but $b_{12} \ne 1$ see
\cite{PE:Bor2003JCP}; for a numerical example see
Fig.~\ref{fig:numerical} and the discussion at the end of this
section). It can be proved that, if $b_{12}=1$ and $a_{12}=0$, then
\mbox{$\mathrm{RMSD}(V_{1},V_{2})=d_{12}/{\sqrt{2}}$}.

\item ER, in turn, only accounts for a systematic error between the
two potentials (an offset). The relation
\mbox{$\mathrm{ER}(V_{1},V_{2})={\mu}_{2}-{\mu}_{1}$} holds and
ER equals the energy-reference shift $a_{12}$
if $b_{12}=1$ (see Eq.~\ref{eq:estimators_b}).
Thus, the changes in the conformational behavior of the system
are not reflected by this quantity.

\item In SDER, the standard deviation associated to ER, the
reference shift is removed by subtracting ER from each
difference ${\Delta}V_{12}^{i}$. However, this quantity
still overestimates the effect of a slope $b_{12} \ne 1$, in fact,
only if $b_{12}=1$, one has that
\mbox{$\mathrm{SDER}(V_{1},V_{2})=d_{12}/{\sqrt{2}}$}.

\item To establish precise relations between AER and $d_{12}$ is
difficult because of the modulus function that enters this
quantity. Nevertheless, it is clear from its definition that AER, like
ER, overestimates both the effect of an energy reference shift
$a_{12}$ and of a slope $b_{12} \ne 1$. For a numerical check of this
fact, see Tab.~\ref{tab:numerical}.

\item Finally, the measure REL, introduced in \cite{PE:Gho1998JPCB},
has much of the spirit of the distance defined in this work. On one
hand, it focuses on the energy differences, which are indeed the
relevant physical quantities to study the conformational behavior of
the system, on the other hand, it correctly removes the effect of an
energy reference shift $a_{12}$. However, it still overestimates the
importance of a slope $b_{12} \ne 1$ and one only has that
\mbox{$\mathrm{REL}(V_{1},V_{2})=d_{12}$} if $b_{12}=1$.
\end{itemize}

There is yet another quantity commonly used for measuring the
differences between two potentials: the Pearson's correlation
coefficient (denoted by $r_{12}$ in the following):

\begin{equation}
\label{eq:r12}
r_{12}:=\frac{\mathrm{Cov}(V_{1},V_{2})}{{\sigma}_{1}{\sigma}_{2}} \  .
\end{equation}

This statistical measure differs from the ones discussed above in
several points. On one hand, it has no units; a fact that renders
difficult to extract from its value relevant statements about the
energies studied. Some statistical statements can be made about the
real value of $r_{12}$ (the value in an infinite sample, denoted by
${\rho}_{12}$), however, to do this, the sampling distribution of
$r_{12}$ must be known. Without making stringent assumptions about the
joint distribution $P_{12}(V_{1},V_{2})$ (see
Sec.~\ref{sec:hypothesis}) only the null hypothesis of ${\rho}_{12}$
being equal to $0$ can be rejected from the knowledge of $r_{12}$ in a
finite sample \cite{PE:Dob1991BOOK}. This is clearly insufficient,
because, in the vast majority of the cases, the researcher {\it knows}
that the two potentials are correlated, i.e., the null hypothesis can
be easily rejected from a priori considerations. If, in turn, one
assumes $P_{12}(V_{1},V_{2})$ to be bivariate Gaussian, the Fisher
transformation can be used to make inferences about ${\rho}_{12}$
which are more general than ${\rho}_{12} \ne 0$
\cite{PE:Dob1991BOOK}. In any case, unfortunately, these type of statements
are not directly translated into statements regarding the energies;
a fact that undermines much of the physical meaning in $r_{12}$.

On the other hand and despite the disadvantages remarked in the
preceding lines, $r_{12}$ behaves satisfactorily when an energy
reference $a_{12}$ is added or when a rescaling $b_{12} \ne 1$ is
introduced between $V_{1}$ and $V_{2}$; like $d_{12}$, and in contrast
to RMSD, ER, SDER, AER and REL, the Pearson's correlation coefficient
does not overestimate such transformations, in fact, $r_{12}$ is
completely insensitive to them. Therefore, it is not surprising that a
simple general relation can be written between both $r_{12}$ and
$d_{12}$:

\begin{subequations}
\label{eq:d12_vs_r12}
\begin{align}
d_{12}&=\sqrt{2}{\sigma}_{2}(1-r_{12}^{2})^{1/2} \  ,
\label{eq:d12_vs_r12_a} \\
d(V_{1},V_{2})&=[({\sigma}_{1}^{2}+{\sigma}_{2}^{2})(1-r_{12}^{2})]^{1/2} \  .
\label{eq:d12_vs_r12_b}
\end{align}
\end{subequations}

In these expressions, it can be observed that the distance herein
introduced depends on two factors: on one side, the width of the
probability distributions associated to the potentials ${\sigma}_{1}$
and ${\sigma}_{2}$, which set the physical scale and give energy units
to $d(V_{1},V_{2})$, on the other, the quantity $1-r_{12}^{2}$, which
measures the degree of correlation between $V_{1}$ and $V_{2}$. The
second factor is completely insensitive to a change in the energy
reference shift or in the slope (due to the properties of $r_{12}$);
the part that depends on the width of the distributions, in turn,
makes the distance sensitive to a change in the slope (remaining
insensitive to a change in the reference), through ${\sigma}_{2}$ if
the rescaling is performed on $V_{2}$ \mbox{(${\sigma}_{2} \rightarrow
{\sigma}_{2}/|b_{12}|$)}. However, contrarily to the case of the
quantities in Eq.~\ref{eq:quantities}, the implications of such a
transformation are not overestimated. In the case of our distance, the
sensitivity to a rescaling arises only from the dilatation of the
random errors, whereas the other quantities take erroneously into
account the fact that the best fit line is not necessarily parallel to
the line $V_{2}=V_{1}$ (see the following numerical example).

In short, {\it the distance defined in this
article consists in a physically meaningful way of giving energy
units to the Pearson's correlation coefficient}.

\begin{table}[!ht]
\begin{tabular}{llrrrrrrr}
\hline \\[-5pt]
& & RMSD & ER & SDER & AER & REL & $r_{12}$ & $d_{12}$ \\[3pt]
\hline \\[-5pt]
{\small $b_{12} \simeq 1$} & {\small $a_{12} \simeq 0$}
& 9.6 & -0.7 & 9.6 & 7.7 & 13.6 & 0.995 & 13.5 \\
{\small $b_{12} \simeq 1$} & {\small $a_{12} \simeq 200$}
& 199.8 & 199.6 & 9.4 & 199.6 & 13.3 & 0.996 & 13.2 \\
{\small $b_{12} \simeq 1/2$} & {\small $a_{12} \simeq 0$}
& 52.7 & -8.4 & 52.0 & 41.9 & 73.8 & 0.980 & 14.4 \\
{\small $b_{12} \simeq 1/2$} & {\small $a_{12} \simeq 200$}
& 205.0 & 198.0 & 52.9 & 198.0 & 74.9 & 0.985 & 13.2 \\[3pt]
\hline \\[-5pt]
\multicolumn{2}{l}{{\small Overestimates $a_{12} \ne 0$}} &
Yes & Yes & No & Yes & No & No & No \\
\multicolumn{2}{l}{{\small Overestimates $b_{12} \ne 1$}} &
Yes & Yes & Yes & Yes & Yes & No & No \\
\multicolumn{2}{l}{{\small Has units of energy}} &
Yes & Yes & Yes & Yes & Yes & No & Yes \\[3pt]
\hline
\end{tabular}
\caption{\label{tab:numerical}{\small Values of the statistical
quantities RMSD, ER, SDER, AER, REL, $r_{12}$ and $d_{12}$
computed in the situations depicted in Fig.~\ref{fig:numerical}.
All the values are in arbitrary energy units except the ones
of $r_{12}$ which have no units. A summary of the properties
of each quantity is presented in the bottom part of the table.}}
\end{table}

To close this section, a numerical example is presented that
summarizes the situations that may be found in practical examples (see
\cite{PE:Bor2003JCP} for a real case of the issues raised) and that
makes explicit the aforementioned disadvantages of the commonly used
statistical quantities. In Fig.~\ref{fig:numerical}, four samples of
$200$ conformations are depicted with the values of $V_{1}$ in the
$x$-axis and the ones of $V_{2}$ in the $y$-axis (both in arbitrary
energy units). The different situations correspond to all generic
cases in which $a_{12}=0$ or $a_{12} \ne 0$ and in which $b_{12}=1$ or
$b_{12} \ne 1$. All the quantities discussed in this section,
including $d_{12}$, have been computed in each case and their values
are presented in Tab.~\ref{tab:numerical}.

From these data and the preceding discussion, some conclusions may be
extracted. First, among the quantities with energy units, SDER and REL
are the most proximate to the distance $d_{12}$, although they will
overestimate the difference between potentials in situations in which
there is a constant rescaling $b_{12} \ne 1$ between them. In
Fig.~\ref{fig:numerical}{\it c}, for example, the contribution of the
points that lie further apart from the origin of coordinates is
overestimated by all the quantities in Eq.~\ref{eq:quantities} for the
sole fact that the best fit-line and the line $V_{2}=V_{1}$ are not
parallel (note that $a_{12}=0$ and that the random noise associated to
these points is not particularly large compared to the one that
corresponds to the points in the central region of the figure). This
is due to the fact that all quantities in Eq.~\ref{eq:quantities} are
based on ${\Delta}V_{12}^{i}$, which measures the distance of each
point to the line $V_{2}=V_{1}$. A disadvantage that is not shared by
$d_{12}$, which measures the differences with respect to the best-fit
line.

Second, the Pearson's correlation coefficient $r_{12}$ has good
properties, although no physically relevant statements can be
extracted from its value due, among other reasons, to the fact that it
does not have units. In Tab.~\ref{tab:numerical}, for example, the
value of $r_{12}$ is close enough to 1 to be considered as a sound
sign of correlation, however, the value of $d_{12}$ (if we pretend it
to be in kcal/mol, which could be the case) tells us that the typical
indetermination in the energy differences, when substituting $V_{1}$
by $V_{2}$, is around 13 kcal/mol, a value an order of magnitude
larger than $RT$. As explained in Sec.~\ref{sec:values}, this suggests
that the relevant behavior of the system may be essentially modified.

Finally, it is worth stressing that all the considerations made in
this section and throughout the article are valid when the physical
quantities compared are potential energy functions of the same system
or closely related systems (see Sec.~\ref{sec:pos_applications}). When
other quantities, such as the $pK_{a}$, charges, dipoles, Born radii,
etc. or energies of distinct systems are the object of the comparison,
the assessment of the discrepancies rests on different theoretical
basis and, frequently, only semi-quantitative statements can be
made. Acknowledging this limitation, the use of any of the quantities
studied in this section, {\it including} $d_{12}$, may be fully justified.
Note, in addition, that the numerical effort needed for the calculation
of $d_{12}$ is both low and very similar to the one required to
compute any of the other quantities (see Sec.~\ref{sec:conclusions}).

\section{Additivity}
\label{sec:additivity}

Frequently, the potentials compared are instances of only a part
of the total potential energy of the system. If the conclusions
extracted, via $d(V_{1},V_{2})$, in such a case are pretended to
be meaningfully transferred to the total energy, this measure
of the difference between potentials must obey some reasonable
additivity rules. Here, we will see that, for some relevant cases
in which certain independence hypothesis are fulfilled, our
distance is approximately additive, although, in other
relevant situations, it is not.

For the sake of brevity, the notation will be much relaxed in
this section and we will assume that we are working with six
different potentials, $x$, $y$, $p$, $r$, $q$ and $s$, that
satisfy the following relations:

\begin{equation}
\label{eq:xpq_yrs}
x=p+q \qquad \mathrm{and} \qquad y=r+s \  .
\end{equation}

Conceptually, $x$ and $y$ must be regarded as instances of the same
potential energy and the same can be said about the pair $p$ and $r$
and the pair $q$ and $s$. Hence, the study of the additivity of our
distance rests on finding a way of expressing $d(x,y)$ as a function
of $d(p,r)$ and $d(q,s)$. If one assumes that $p$ is independent from
$q$ and that $r$ is independent from $s$ (see the discussion at the
end of this section for the implications of such an hypothesis), one
has that $r_{pq}=0$, $r_{rs}=0$,
${\sigma}_{x}^{2}={\sigma}_{p}^{2}+{\sigma}_{q}^{2}$ and
${\sigma}_{y}^{2}={\sigma}_{r}^{2}+{\sigma}_{s}^{2}$. In such a case,
the following additivity relation can be written:

\begin{equation}
\label{eq:additivity}
d^{2}(x,y)=d^{2}(p,r)+d^{2}(q,s)+{\Delta}d \  ,
\end{equation}

where:

\begin{equation}
\label{eq:def_Dd}
{\Delta}d:=({\sigma}_{p}^{2}+{\sigma}_{r}^{2})(r_{pr}^{2}-r_{xy}^{2})+
           ({\sigma}_{q}^{2}+{\sigma}_{s}^{2})(r_{qs}^{2}-r_{xy}^{2}) \  .
\end{equation}

And the correlation coefficient $r_{xy}$ can be expressed in terms
of quantities associated to $p$, $r$, $q$ and $s$ in the
following way (note that $r_{xy}$ is indeed not additive):

\begin{equation}
\label{eq:rxy}
r_{xy}=\frac{{\sigma}_{p}{\sigma}_{r}r_{pr}+{\sigma}_{q}{\sigma}_{s}r_{qs}}
            {({\sigma}_{p}^{2}+{\sigma}_{q}^{2})^{1/2}
             ({\sigma}_{r}^{2}+{\sigma}_{s}^{2})^{1/2}} \  .
\end{equation}

Now, one can see in Eq.~\ref{eq:additivity} that, if ${\Delta}d$ were
zero, the square of the distance would be exactly additive in the
aforementioned sense, making it possible to assert, for example, that,
if $p$ is proximate to $r$ and $q$ is proximate to $s$, then $x=p+q$
is proximate to $y=r+s$. Unfortunately, this is not the case. It can
be shown that ${\Delta}d \ge 0$ (the distance is {\it over-additive})
and, without imposing any restriction on the
potentials studied, nothing satisfactory can be said. For example, a
particularly undesirable, albeit also uncommon, situation is that for
which $\mathrm{Cov}(p,r)=-\mathrm{Cov}(q,s)$. Such a relation, makes
zero the numerator in Eq.~\ref{eq:rxy} and, consequently, $r_{xy}$.
Substituting $r_{xy}=0$ in Eq.~\ref{eq:def_Dd} and taking ${\Delta}d$
to Eq.~\ref{eq:additivity}, one has that, for every allowed value of
$r_{pr}$ and $r_{qs}$:

\begin{equation}
\label{eq:add_worst}
\mathrm{Cov}(p,r)=-\mathrm{Cov}(q,s) \Rightarrow
d^{2}(x,y)={\sigma}_{p}^{2}+{\sigma}_{q}^{2}+{\sigma}_{r}^{2}+{\sigma}_{s}^{2}=
{\sigma}_{x}^{2}+{\sigma}_{y}^{2} \  ,
\end{equation}

which is the worst possible value of $d(x,y)$.

However, there exists
a particular class of situations than can be argued to be proximate
to the situations found in typical cases and for which
the additivity is approximately accomplished. These special situations
are characterized for the satisfaction of the following relation:

\begin{equation}
\label{eq:k}
{\sigma}_{p}/{\sigma}_{r}={\sigma}_{q}/{\sigma}_{s}:=k \  .
\end{equation}

When this equality is satisfied, it can be proved that the following
quotient:

\begin{equation}
\label{eq:def_Dd_rel}
{\Delta}d_{\mathrm{rel}}:={\Delta}d/(d^{2}(p,r)+d^{2}(q,s)) \  ,
\end{equation}

which measures the relative deviation from the exact additivity, does
not depend on $k$ and can be expressed as a function of only
${\sigma}_{r}$, ${\sigma}_{s}$, $r_{pr}$ and $r_{qs}$. If, in
addition, we define $c$ through ${\sigma}_{s}=c\, {\sigma}_{r}$,
without loss of generality, we can write ${\Delta}d_{\mathrm{rel}}$
as a function of only $r_{pr}$, $r_{qs}$, and $c$ as follows:

\begin{equation}
\label{eq:Dd_rel}
{\Delta}d_{\mathrm{rel}}=\frac{c^{2}(r_{pr}-r_{qs})^{2}}
  {(1+c^{2})(1-r_{pr}^{2}+c^{2}(1-r_{qs}^{2}))} \  .
\end{equation}

\begin{figure}[!ht]
\begin{center}
\epsfig{file=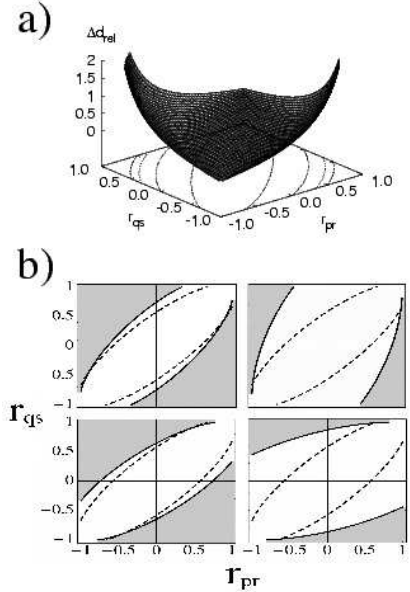,width=5.5cm}
\end{center}
\caption{\label{fig:additivity}{\small Graphical study of the
additivity of the distance. {\bf (a)} ${\Delta}d_{\mathrm{rel}}$ as a
function of $r_{pr}$ and $r_{qs}$ for $c=1$ (see Eq.~\ref{eq:Dd_rel}).
Contour lines are plotted at the levels
${\Delta}d_{\mathrm{rel}}=10\% ,50\% ,100\% ,150\%$. {\bf (b)} In white, the
regions in $(r_{pr},r_{qs})$-space with ${\Delta}d_{\mathrm{rel}}<10\%$
for different values of $c$. From left to right and from top to
bottom, each figure corresponds to $c=1/2,1/5,2,5$. In each
case, the borders of the ${\Delta}d_{\mathrm{rel}}<10\%$ region
for $c=1$ are shown with broken lines for comparison. }}
\end{figure}

Representing this equation as a three-dimensional surface (see
Fig.~\ref{fig:additivity}{\it a}), one has a {\it valley} whose lowest
region lies in the line $r_{pr}=r_{qs}$ and has zero height, i.e.,
\mbox{${\Delta}d_{\mathrm{rel}}(r_{pr}=r_{qs})=0$}. The slopes of the
valley are curved and ascend as one moves away from the minimum line,
eventually reaching arbitrarily large values of
${\Delta}d_{\mathrm{rel}}$ when $(r_{pr},r_{qs}) \rightarrow (1,-1)$
or $(r_{pr},r_{qs}) \rightarrow (-1,1)$.

Numerically, the region for which the value of
${\Delta}d_{\mathrm{rel}}$ is acceptable is rather large. In
Fig.~\ref{fig:additivity}{\it b}, the contour lines corresponding to
${\Delta}d_{\mathrm{rel}}=10\%$ are depicted for some values of $c$
that may be found in practical cases. It can be seen that, as one
departs from \mbox{$c=1$}, the region for which
${\Delta}d_{\mathrm{rel}}<10\%$ gets larger, occupying, in any case,
the majority of the $(r_{pr},r_{qs})$-space. Therefore, one can
conclude that, for the cases in which Eq.~\ref{eq:k} is satisfied, the
square of the distance introduced in this article is approximately
additive in the relevant situations in which the correlations between
$p$ and $r$ and between $q$ and $s$ are similar. Moreover, for
continuity arguments, one has that, in the case that Eq.~\ref{eq:k}
were only approximately satisfied, the situation would be proximate to
the one described in the previous lines.

Finally, some remarks must be made about the assumption of
independence between $p$ and $q$ and between $r$ and $s$. At first
sight, one would say that this hypothesis, as the independence
hypothesis in Sec.~\ref{sec:hypothesis}, is under researcher's
control. In the case of a generic complex system (a spin glass, a
random heteropolymer, etc.), this is indeed the case, however, if the
object of study is a protein, one must be cautious. It is widely
believed that the sequences of proteins are the result of a
million-years-long selection process whose driving force is the search
for the ability to fold rapidly and robustly
\cite{PE:Alo2004BOOK,PE:Onu2000APC,PE:Dil1999PSC,PE:Dob1998ACIE}. Regarding
the interactions responsible of the folding process, this means that
they have been optimized in the sequence space to be {\it minimally
frustrated} \cite{PE:Bry1987PNAS}, i.e., maximally cooperative. In
such a case, the correlations between different parts of the total
potential energy may be large and the study of the additivity done in
this section should be regarded as a privileged reference situation.

\section{Application}
\label{sec:application}

\begin{figure}[!ht]
\begin{center}
\epsfig{file=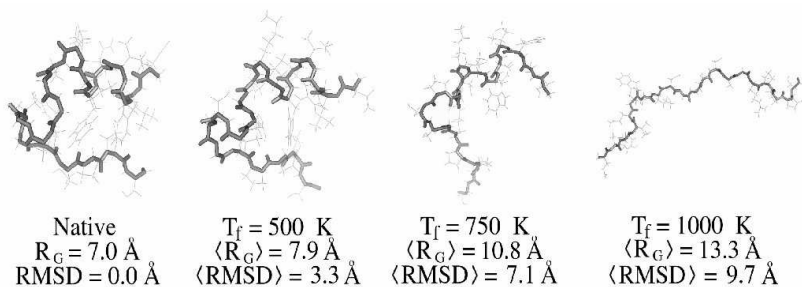,width=11cm}
\end{center}
\caption{\label{fig:confs}{\small Native conformation of the Trp-Cage
protein together with arbitrarily chosen structures from three
particular subsets of the working set. The average radius of gyration
$\langle R_{G} \rangle$ and the average RMSD with respect to the
native structure is presented for each set. Both quantities have been
computed taking into account only the $\alpha$-carbons. Pictures
generated with {\tt PyMOL} (DeLano, W. L., 2002, {\tt
http://www.pymol.org}). }}
\end{figure}

To illustrate one of the possible practical applications of the
distance, we first study the robustness of the van der Waals energy, as
implemented in the CHARMM molecular dynamics program
\cite{PE:Mac1998BOOK,PE:Bro1983JCC}, in a particular system: the de
novo designed protein known as \mbox{Trp-Cage} \cite{PE:Nei2002NSB}
(PDB code 1L2Y).

The program CHARMM itself was used as a conformation
generator. From the native conformation stored in the Protein Data
Bank \cite{PE:Ber2000NAR}, a 10 ps heating dynamics\footnote{ The
{\tt c27b4} version of the CHARMM program was used. The molecular
dynamics were performed using the {\it Leap Frog} algorithm therein
implemented and the {\tt param22} parameter set, which is optimized
for proteins and nucleic acids. The water was taken into account
implicitly with the \cite{PE:Dom1999JPCB} version of the Generalized
Born Model built into the program. } was performed on the system, from
an initial temperature \mbox{$T_{i}=0$ K} to eleven different final
temperatures (from $T_{f}=500$ K to $T_{f}=1000$ K in steps of 50
K). This was repeated 100 times for each final temperature with a
different seed for the random numbers generator each time. The overall
result of the process was the production of a working set of 1100
different conformations of the protein, whose structures range from
{\it close to native} (the $T_{f}=500$ K set) to {\it completely
unfolded} (the $T_{f}=1000$ K set) (see Fig.~\ref{fig:confs}). It is
worth remarking that the short time in which the system was heated (10
ps) and the fact that there was no equilibration after this process
cause the final temperatures to be only {\it labels} for the eleven
aforementioned sets of conformations. They are, by no means, the
thermodynamical temperatures of any equilibrium state from which the
structures are taken. This sets of conformations are only meant to
sample the representative regions of the phase space. In
Fig.~\ref{fig:confs}, arbitrarily chosen structures from three
particular sets are shown together with the native conformation. The
average radius of gyration $\langle R_{G} \rangle$ of each set,
depicted in the same figure, must be compared to the radius of
gyration of the native state\footnote{\label{foot:ca}Both $R_{G}$ and
the RMSD have been computed taking into account only the
$\alpha$-carbons.}. The average RMSD of the structures in each set
with respect to the native structure, calculated via the
quaternion-based method described in \cite{PE:Cou2004JCC}, is also
presented\footnote{\label{foot:RMSD}The notation for this quantity,
which is the root mean square deviation of the atomic coordinates of
two structures after optimal superposition \cite{PE:Cou2004JCC}, is
the same as the one used for the RMSD of the energies in
Sec.~\ref{sec:relation}. This choice has been made for consistency
with the literature, in which this ambiguity is very common.}.

The van der Waals energy implemented in CHARMM may be expressed
as follows:

\begin{equation}
\label{eq:vdW}
V:=\sum_{i<j} \left [ ({\varepsilon}_{i}{\varepsilon}_{j})^{1/2}
  \left ( \left ( \frac{R_{i}+R_{j}}{r_{ij}} \right )^{12} -
  2 \left ( \frac{R_{i}+R_{j}}{r_{ij}} \right )^{6} \right ) \right ] \  ,
\end{equation}

where the sum is extended to all the pairs of atoms and the free
parameters ${\varepsilon}_{i}$ and $R_{i}$ only depend on the type of
atom (i.e., two atoms $i$ and $j$ of the same type have the same
parameters assigned).

Using the working set of conformations described above of the Trp-Cage
protein, the robustness of this potential energy function to changes
in the free parameters ${\varepsilon}_{C}$ and $R_{C}$ associated to
the aliphatic sp3 carbon CH (denoted by CT1 in CHARMM) is
investigated. To do this, a finite grid-like set of points
$({\varepsilon}_{C}^{k},R_{C}^{k})$ is chosen in the bi-dimensional
parameter space, with ${\varepsilon}_{C}^{k}$ ranging from
\mbox{$-0.10$ kcal/mol} to \mbox{$-0.02$ kcal/mol} and $R_{C}^{k}$
ranging from \mbox{$2$ \AA} to \mbox{$4$ \AA}. Then, for each point in
this set, different values ${\delta}{\varepsilon}_{C}$ are added to
and subtracted from ${\varepsilon}_{C}^{k}$, or different values
${\delta}R_{C}$ are added to and subtracted from $R_{C}^{k}$
independently. The potential that corresponds to
\mbox{${\varepsilon}_{C}={\varepsilon}_{C}^{k}-{\delta}{\varepsilon}_{C}$}
is denoted by $V_{1}$, the one that corresponds to
\mbox{${\varepsilon}_{C}={\varepsilon}_{C}^{k}+{\delta}{\varepsilon}_{C}$}
is denoted by $V_{2}$ (analogously with $R_{C}$) and the distance
$d(V_{1},V_{2})$ between the two instances is computed in each case
(i.e., for each {\it central point}
$({\varepsilon}_{C}^{k},R_{C}^{k})$ and for each
${\delta}{\varepsilon}_{C}$ (or ${\delta}R_{C}$))\footnote{
\label{foot:hyp_fulfilled} It can be proved that, in this particular
case, the normality hypothesis in eq.~\ref{eq:cond_pdfs_V1V2} is
approximately fulfilled.}.

\begin{figure}[!ht]
\begin{center}
\epsfig{file=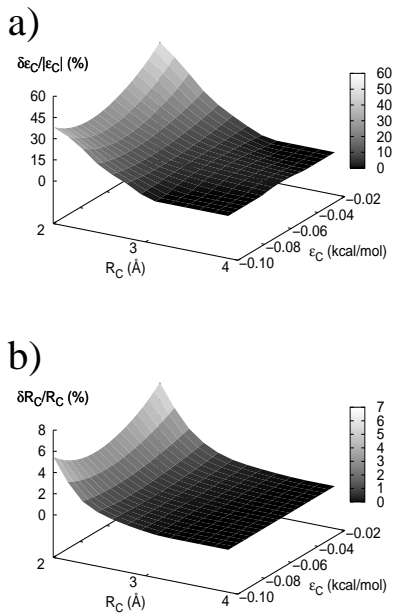,width=5.5cm}
\end{center}
\caption{\label{fig:robustness}{\small Robustness of the van der Waals
energy to changes in some free parameters. Relative indetermination in
${\varepsilon}_{C}$ {\bf (a)} and in $R_{C}$ {\bf (b)} associated to
$d(V_{1},V_{2})=RT$ as a function of the central point in the
parameter space. Larger values of the relative indetermination
correspond to greater robustness. }}
\end{figure}

This procedure allows us to study the dependence of the distance
between $V_{1}$ and $V_{2}$ on the corresponding difference,
${\delta}{\varepsilon}_{C}$ or ${\delta}R_{C}$, between this two
potentials in the parameter space for each central point
$({\varepsilon}_{C}^{k},R_{C}^{k})$. This relation may be regarded as
one between indetermination in the values of the free parameters and
its influence on the conformational behavior of the system. From this
point of view, the difference ${\delta}{\varepsilon}_{C}$ (or
${\delta}R_{C}$) in the parameter space for which the distance
associated equals $RT$ (see Sec.~\ref{sec:values}) must be considered
an amount of indetermination in the parameters that does not involve
relevant physical changes in the system. Therefore, if the parameters
are known to a precision equal or greater than the one associated to
these particular values of ${\delta}{\varepsilon}_{C}$ or
${\delta}R_{C}$, the statistical indetermination of the parameters may
be regarded as harmless. The values of this differences (as a function
of the central point $({\varepsilon}_{C}^{k},R_{C}^{k})$) computed for
the system studied in this section are depicted in
Fig.~\ref{fig:robustness}.

Although this study only pretends to be an illustration of the
concepts introduced in the previous sections and more features of the
van der Waals energy should be investigated elsewhere, some
interesting remarks may be made about the results herein
presented. One one hand, directly from Fig.~\ref{fig:robustness}, one
can see that the precision needed in $R_{C}$ is much greater than the
one needed in ${\varepsilon}_{C}$, i.e., the van der Waals energy is
more sensitive to changes in $R_{C}$ than in ${\varepsilon}_{C}$. This
is reasonable because $V$ depends on $R_{C}$ raised to the 12th and
6th power whereas ${\varepsilon}_{C}$ only enters the expression
raised to $1/2$ (see Eq.~\ref{eq:vdW}). On the other hand, the allowed
indetermination in the parameters grows, in both cases, as $R_{C}$
diminishes (the dependence on ${\varepsilon}_{C}$ is much weaker). The
reason for this being probably that, when the van der Waals radius
$R_{C}$ is large enough, the atoms begin to clash, i.e., the 12th
power in Eq.~\ref{eq:vdW}, associated to the steric repulsion,
begins to dominate over the 6th power term, associated to the
attractive dispersion forces.

Finally, we would like to mention that, for the values
\mbox{${\varepsilon}_{C}=-0.02$ kcal/mol} and \mbox{$R_{C}=2.275$
\AA}, which are the ones used in the CHARMM {\tt param22} parameter
file, the allowed indeterminations in the parameters are
\mbox{${\delta}{\varepsilon}_{C}/|{\varepsilon}_{C}|=35$ \%} and
\mbox{${\delta}R_{C}/R_{C}=3$ \%}, in the region of relatively
lower required precision (i.e., the relatively more favorable
region). Note, however, that the indetermination for $R_{C}$
corresponds to \mbox{$\sim 0.07$ \AA}, which is a rather demanding
accuracy and suggests that, if the van der Waals radii set is changed,
the behavior of the system may be significantly modified.

\begin{figure}[!ht]
\begin{center}
\epsfig{file=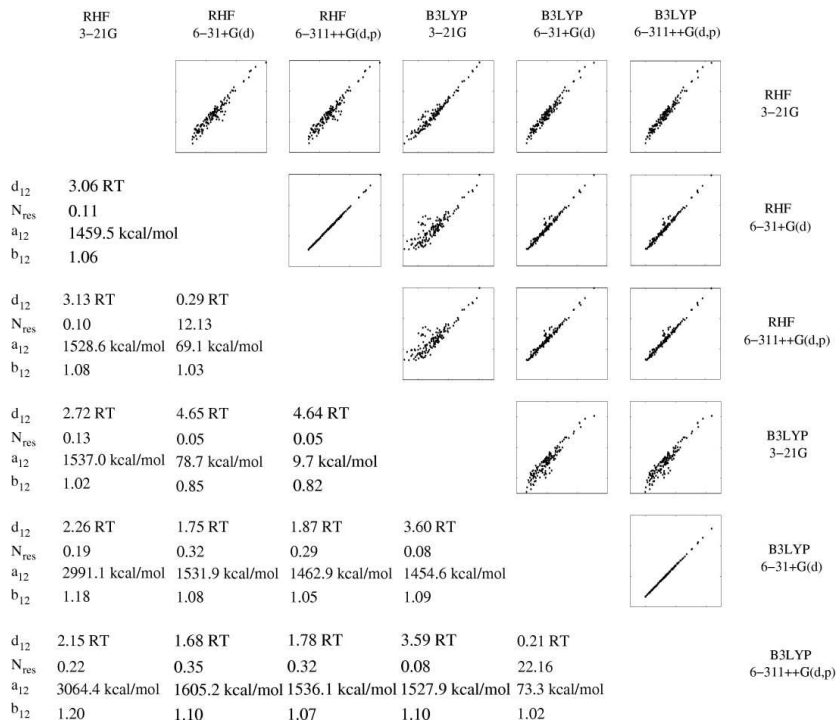,width=11cm}
\end{center}
\caption{\label{fig:for_ala_ami}{\small Comparison between different
levels of the theory in the quantum mechanical ab initio study of the
Potential Energy Surface (PES) associated with the Ramachandran angles
of the model dipeptide HCO-L-Ala-NH$_{2}$.
The figure must be read as follows: {\bf 1)} Any numerical set
of measures is associated to the comparison between the level of the
theory in the corresponding {\it row} (denoted by $V_{1}$) and
{\it column} (denoted by $V_{2}$). {\bf 2)} The
conformations scatter plot that belongs to a particular set of measures is the
one that lies in the position which is obtained via reflection (of the set)
with respect to the blank diagonal.}}
\end{figure}

As a second brief example of the possible applications of the method,
we present an exploratory comparison of different levels of theory
in the quantum mechanical ab initio study of the Potential Energy
Surface (PES) associated with the Ramachandran angles of the
model dipeptide HCO-L-Ala-NH$_{2}$. This comparison is an example
of the first point discussed in Sec.~\ref{sec:pos_applications}.

In \cite{PE:Per2003JCC},
the PES of HCO-L-Ala-NH$_{2}$ is calculated with two methods,
RHF and B3LYP, using, for each one, three different basis sets,
3-21G, 6-31+G(d) and 6-311++G(d,p). To do this, the Ramachandran
space is divided in a 12x12 grid and, fixing the values of the
$\Phi$ and $\Psi$ torsional angles, a geometrical optimization
of the structure is performed at each point. This process
produces the values of six different instances of the same
potential energy on a working set of 144 conformations of the
system.

In Fig.~\ref{fig:for_ala_ami}, each one of the six levels of the
theory is compared to the other five (using the data provided by
A. Perczel) and some relevant numerical
measures are presented. The distance $d_{12}$ is given in units of
$RT$ (at 300 K), the fitted energy reference shift $a_{12}$ and slope
$b_{12}$ are also shown and the only quantity that requires further
explanation is $N_{\mathrm{res}}$ (see Eq.~\ref{eq:N_res} below).

One of the interests in studying PESs of peptide models lies on the
possibility of using the results for modeling short oligo-peptides or
even proteins \cite{PE:Alo2004BOOK}. If we imagine that we use the PES
of HCO-L-Ala-NH$_{2}$ for constructing a potential that describes the
behavior of a peptide formed by $N$ alanine residues, the first {\it
naive} attempt would be to simply add $N$ times the potential energy
surface of the individual HCO-L-Ala-NH$_{2}$ (making each term
suitably depend on different pairs of Ramachandran angles). We may now
ask whether the distance between two different instances of the
$N$-residue peptide potential can be related to the distance between
the corresponding mono-residue ones. It can be proved, appropriately
choosing the working set of conformations of the larger system and
using the relations presented in Sec.~\ref{sec:additivity}, that the
following relation holds:

\begin{equation}
\label{eq:d_N}
d_{12}(N) = \sqrt{N} d_{12}(1) \  ,
\end{equation}

where we have denoted by $d_{12}(N)$ the distance
between the $V_{1}$ and $V_{2}$ potentials of the $N$-residue peptide.

Hence, we define $N_{\mathrm{res}}$ as the $N$ for which
$d_{12}(N)=RT$, representing up to which number of residues
the criterium given in Sec.~\ref{sec:values} will be satisfied:

\begin{equation}
\label{eq:N_res}
d_{12}(N_{\mathrm{res}}) := RT \quad \Longrightarrow \quad
  N_{\mathrm{res}} = \left ( \frac{RT}{d_{12}(1)} \right )^{2} \  .
\end{equation}

Although a much more exhaustive study will be carried out in future
works, let us extract some meaningful conclusions from the data in
Fig.~\ref{fig:for_ala_ami}. Note, first, that the only two cases for
which $d_{12} < RT$ are RHF/6-31+G(d) vs. RHF/6-311++G(d,p) and
B3LYP/6-31+G(d) vs. B3LYP/6-311++G(d,p). This means that the
convergence in basis is achieved for both methods somewhere between
6-31+G(d) and 6-311++G(d,p) and suggests (for HCO-L-Ala-NH$_{2}$) that
there is no need in going above 6-31+G(d). Of course, the fact that
$N_{\mathrm{res}} \simeq 22$, in the B3LYP case, and
$N_{\mathrm{res}} \simeq 12$, in the RHF case, places a limit on the size
of the system for which the similarity of the two levels should be
considered as sufficient. Finally, note that the distance between
RHF/6-311++G(d,p) and B3LYP/6-311++G(d,p) is 1.78 RT, which means that
the convergence in methods has not been achieved and some more
accurate method should be studied.

\section{Conclusions}
\label{sec:conclusions}

In this work, a measure $d(V_{1},V_{2})$ of the differences between two
instances of the same potential energy has been defined and
the following points about it have been discussed:

\begin{itemize}
\item It rests on hypothesis whose validity stems from general
characteristics shared by many complex systems and from the
statistical laws of large numbers. We believe that, without knowing
specific details of the system, the statistical approach is
unavoidable and, among the many criteria, our distance is
the most meaningful way of quantifying the differences
between potentials.
\item It allows to make physically meaningful statements about the way
in which the energy differences between conformations change (or how
the energetic ordering of the conformations is altered) upon
substitution of one potential by the other.
\item It may be applied to at least three practical situations
characterized by the origin of the differences between the potentials.
\begin{itemize}
\item Different algorithms or approximations are used (potential design).
\item The potential energy function depends on free parameters and the
two instances correspond to different values of them (robustness).
\item Slightly different systems are compared (mutational studies,
effective potentials).
\end{itemize}
\item It presents advantages over the commonly used quantities RMSD, ER,
SDER and AER that consist mainly of not overestimating irrelevant
transformations on the potentials, such as adding an energy reference
or rescaling one of them. Regarding the Pearson's correlation coefficient
$r$, our distance may be considered as a physically meaningful way of
giving him energy units. Finally, the numerical complexity involved
in the calculation of $d(V_{1},V_{2})$ (see below) is similar to the one
associated to any of the other quantities.
\item It is approximately additive for most of the interesting situations
encountered in practical cases.
\end{itemize}

In addition, a first practical example, which consists in the study of
the robustness to changes in the free parameters of the van der Waals
energy in CHARMM, and a second one, in which the ab initio potential
energy surfaces of the HCO-L-Ala-NH$_{2}$ molecule calculated at
different levels of the theory are compared, have been presented to
illustrate the concepts discussed.

Finally, we would like to summarize the steps that must be followed to
compute the distance in a practical case. Although all that follows
has been said, we believe that a brief {\it recipe} could be useful
for quick reference:

\begin{enumerate}
\item Generate a working set of independent conformations $\{{\vec
q}_{i}\}^{N}_{i=1}$ (see Sec.~\ref{sec:hypothesis} and the last
paragraph of Sec.~\ref{sec:definition}).
\item Denote $V_{1}^{i}:=V_{1}({\vec q}_{i})$, $V_{2}^{i}:=V_{2}({\vec
q}_{i})$ and compute the statistical quantities ${\mu}_{1}$,
${\mu}_{2}$, ${\sigma}_{1}$ and $\mathrm{Cov}(V_{1},V_{2})$ in
Eq.~\ref{eq:stat_quantities}.
\item With them, calculate the mean-square estimators through
Eq.~\ref{eq:estimators}. First $b_{12}$, then $a_{12}$ and, finally
${\sigma}_{12}$.
\item If comparing a potential to a more accurate instance, use
$d_{12}=\sqrt{2}{\sigma}_{12}$ to find the asymmetrical version of the
distance between them, and rescale $V_{2}$ dividing it by
$b_{12}$ if desired. Otherwise, repeat the steps 2 and 3 changing $1
\leftrightarrow 2$ in all the expressions to compute ${\sigma}_{21}$
and use Eq.~\ref{eq:def_d} to finally arrive to $d(V_{1},V_{2})$.
\item If $d(V_{1},V_{2})<RT$ (or $d_{12}<RT$, depending on the case),
the two potentials may be considered physically equivalent.
\end{enumerate}

\section*{Appendix A: Metric properties}

For completeness, and because, in the case of our distance, it is
illustrative to do so, we will investigate, in the following, in which
situations (which will turn out to be rather common) the behavior of
$d(V_{1},V_{2})$ approaches that of a traditional mathematical {\it
distance}. However, it must be stressed that the measure introduced in
this article was never intended to be such an object. Its meaning is
encoded in the statistical statements derived from its value (see
Sec.~\ref{sec:meaning}) and the name {\it distance} must be used in a
more relaxed manner than the one traditionally found in mathematics.

The object $\mathcal{D}(x,y)$ is said to be a {\it distance} (also a
{\it metric}) in mathematics if it satisfies the following properties:

\begin{enumerate}
\item $\mathcal{D}(x,y)=0 \Leftrightarrow x=y$
\item Positivity: $\mathcal{D}(x,y) \ge 0$
\item Symmetry: $\mathcal{D}(x,y)=\mathcal{D}(y,x)$
\item Triangle inequality: $\mathcal{D}(x,z) \le
\mathcal{D}(x,y)+\mathcal{D}(y,z)$
\end{enumerate}

Whereas, in the case of $d(V_{1},V_{2})$:

\begin{enumerate}
\item The first property is not fulfilled. One
certainly has the implication to the left, but the direct implication
is false in general. As has been stated in the Pt.~\ref{point:linear}, in
Sec.~\ref{sec:meaning}, the analogous property that $d(V_{1},V_{2})$
satisfies is that $d(V_{1},V_{2})=0$ is equivalent to $V_{2}$ being
a linear transformation of $V_{1}$ and vice versa, i.e., to
$V_{2}({\vec q}_{i})=b_{12}V_{1}({\vec q}_{i})+a_{12}$ and 
$V_{1}({\vec q}_{i})=b_{21}V_{2}({\vec q}_{i})+a_{21}$,
$\forall {\vec q}_{i} \in \{{\vec q}_{i}\}^{N}_{i=1}$. Where,
additionally, one has that $b_{21}=1/b_{12}$ and $a_{21}=-a_{12}/b_{12}$.
The fact that this property of a mathematical distance is not
satisfied by $d(V_{1},V_{2})$ must be considered an advantage,
because, as has been remarked in previous sections, it is reasonable
to regard as equivalent two potentials if there is only a linear
transformation between them.

\item $d(V_{1},V_{2}) \ge 0$ for every $V_{1}$ and $V_{2}$.

\item Directly from its definition in Eq.~\ref{eq:def_d} or from
Eq.~\ref{eq:d12}, it is evident that $d(V_{1},V_{2})$ is symmetrical
under change of $V_{1}$ by $V_{2}$. This property is not fulfilled by
the quantity $d_{12}$. However, the situation in which it is
reasonable to use it (the comparison of a particular instance of a
potential energy $V$ to a less accurate one) is also intrinsically
asymmetrical (see the final part of Sec.~\ref{sec:meaning}).

\item The triangle inequality, in this context, is a relation that
must be expressed as a function of the statistical quantities related
to three different potentials, $V_{1}$, $V_{2}$ and $V_{3}$, as
follows:
\begin{equation}
\label{eq:triangle}
\sqrt{{\sigma}_{1}^{2}+{\sigma}_{3}^{2}}\sqrt{1-r_{13}^{2}} \le
\sqrt{{\sigma}_{1}^{2}+{\sigma}_{2}^{2}}\sqrt{1-r_{12}^{2}} +
\sqrt{{\sigma}_{2}^{2}+{\sigma}_{3}^{2}}\sqrt{1-r_{23}^{2}} \  .
\end{equation}
This relation is not fulfilled for every triplet
$(V_{1},V_{2},V_{3})$, i.e., the distance introduced in this article
does not satisfy, in general, the triangle inequality. A simple
counterexample is found if one makes ${\sigma}_{3}$ grow, keeping the
rest of the quantities in Eq.~\ref{eq:triangle} constant. For
${\sigma}_{3}$ large enough, the relation above may be approximated by:
\begin{equation}
\label{eq:triangle_ap}
{\sigma}_{3} \left [ \sqrt{1-r_{13}^{2}} - \sqrt{1-r_{23}^{2}} \right ] \le
\sqrt{{\sigma}_{1}^{2}+{\sigma}_{2}^{2}}\sqrt{1-r_{12}^{2}} \  .
\end{equation}
Then, if $r_{13}^{2} < r_{23}^{2}$, one may make ${\sigma}_{3}$ even
larger and eventually break the inequality (in the case that it were
not already broken for the value of ${\sigma}_{3}$ for which
Eq.~\ref{eq:triangle_ap} is a good approximation). As a final
remark, it is worth pointing out that, despite the general
mathematical facts stated above, there is a particular situation,
which is also expected to be similar to the situations
relevant to be studied, in which the distance has been found
to satisfy the triangle inequality. If one has that
${\sigma}_{1}={\sigma}_{2}={\sigma}_{3}$ (something that
is expected to be approximately true in the case that the three
potentials are proximate), Eq.~\ref{eq:triangle} turns into
a relation involving only the correlation coefficients:
\begin{equation}
\label{eq:triangle_r}
\sqrt{1-r_{13}^{2}} \le \sqrt{1-r_{12}^{2}} + \sqrt{1-r_{23}^{2}} \  .
\end{equation}
In addition, assuming the hypothesis discussed in Sec.~\ref{sec:hypothesis},
the following inequalities can be proved \cite{PE:Dob1991BOOK} without
any further assumptions about the potentials:
\begin{subequations}
\label{eq:rs}
\begin{align}
r_{13} \ge r_{12} r_{23} - \sqrt{1-r_{12}^{2}} \sqrt{1-r_{23}^{2}} \  ,
\label{eq:rs_a} \\
r_{13} \le r_{12} r_{23} + \sqrt{1-r_{12}^{2}} \sqrt{1-r_{23}^{2}} \  .
\label{eq:rs_b}
\end{align}
\end{subequations}
We have numerically found that, if the relations in Eq.~\ref{eq:rs}
are satisfied, so is the one in Eq.~\ref{eq:triangle_r}. Hence, if
${\sigma}_{1}={\sigma}_{2}={\sigma}_{3}$, then, for all values of
$r_{12}$, $r_{23}$ and $r_{13}$, the distance satisfies the triangle
inequality. Clearly, for continuity, if
${\sigma}_{1}={\sigma}_{2}={\sigma}_{3}$ is not exactly but approximately
satisfied, then, although the triangle inequality may be broken,
it will broken by a small relative amount. In short, if one has
${\sigma}_{1}={\sigma}_{2}={\sigma}_{3}$ approximately, one has
the triangle inequality also approximately.
\end{enumerate}

\vspace{0.2cm} {\small We would like to thank I. Calvo, F. Falceto,
A. Jaramillo, V. Laliena, F. Plo and D. Zueco for illuminating
discussions and also to Andr\'as Perczel for providing us with the
data used in the second example of Sec.~\ref{sec:application} and in
Fig.~\ref{fig:v_const}. }

\vspace{0.2cm} {\small This work has been supported by the Arag\'on
Government (``Biocomputaci\'on y F\'{\i}sica de Sistemas Complejos''
group) and by the research grants MEC (Spain) \mbox{FIS2004-05073} and
MCYT (Spain) \mbox{BFM2003-08532}. P. Echenique is supported by a MEC
(Spain) FPU grant.}

\bibliography{distance}

\end{document}